\documentclass [a4paper,fleqn, 12pt]{article}
\usepackage{graphicx}

\usepackage {amsmath} \usepackage{amssymb} \usepackage{cite}

\newcommand{\cn}

\begin{document}


\title
{Exact solutions of equations for the Burgers hierarchy}

\author
{Nikolay A. Kudryashov, \and Dmitry I. Sinelshchikov}

\date{Department of Applied Mathematics, National Research Nuclear University
MEPHI, 31 Kashirskoe Shosse,
115409 Moscow, Russian Federation}




\maketitle

\begin{abstract}
Some classes of the rational, periodic and solitary wave solutions for the Burgers hierarchy are presented. The solutions for this hierarchy are obtained by using the generalized Cole - Hopf transformation.

\end{abstract}






\section{Introduction}

The  Burgers hierarchy is well known family of
nonlinear evolution equations. This hierarchy can be
written in the form
\begin{equation}
\label{BH}
u_t+\alpha\,\frac{\partial}{\partial x}\left(\frac{\partial}{\partial x}+u\right)^{n}u=0,\quad n=0,1,2,\ldots,
\end{equation}
At $n=1$ Eq. \eqref{BH} is the Burgers equation
\begin{equation}
\label{B}
u_t+2\alpha\,u\,u_{x}+\alpha\,u_{xx}=0.
\end{equation}
Eq. \eqref{B} was firstly introduced in \cite{Burgers}. It's well known that the Burgers equation can be linearized by the Cole---Hopf transformation \cite{Hopf51,Cole50}. Exact solutions of Eq.\eqref{B} were discussed in many papers( see for example \cite{Rosenblatt68,Benton66,Malfliet,Fahmy} ).

In the case $n=2$ from  Eq. \eqref{BH} we have the Sharma - Tasso - Olver (STO) equation
\begin{equation}
\label{S}
u_t+\alpha\,u_{xxx}+3\alpha\,u_{x}^{2}+3\,\alpha\,u\,u_{xx}+3\,\alpha\,u^{2}\,u_{x}=0.
\end{equation}
The STO equation was derived in \cite{Tasso,Olver}. Some exact solutions of this equation was obtained in \cite{Hereman,Yang,Gudkov,Wazwaz,Shang,Lou,Kudryashov08}.

At $n=3$ and $n=4$ we have the following fourth and fifth order partial differential equations
\begin{equation}\begin{gathered}
\label{FM}
u_{t} +\alpha\,u_{xxxx}  +10\,\alpha\,
u_{x} u_{xx}  +4\,\alpha\,u
u_{xxx} +12\,\alpha\,u   u_{x}^{2}+\\+6\,\alpha\,
  u  ^{2} u_{xx}  +4\alpha\,u^{3}\,u_{x}=0,
\end{gathered}\end{equation}
\begin{equation}\begin{gathered}
\label{FFM}
u_{t} +\alpha\,u_{xxxxx}  +10\,\alpha\,
u_{xx}^{2}+15\,\alpha\, u_{x}
u_{xxx}  +5\,\alpha\,u u_{xxxx}  +15\,\alpha\,
 u_{x}^
{3}+
\\+50\,\alpha\,u u_{x} u_{xx}  +10\,\alpha\, u^{2}u_{xxx}
  +30\,\alpha\,  u^{2} u_{x}^{2}+10\,\alpha\,u^{3} u_{xx}  +5\,\alpha\, u ^{4}u_{x} =0.
\end{gathered}\end{equation}

In this paper we present the generalized Cole--- Hopf transformation which we use for finding different types of exact solutions: the solitary wave solutions, the periodic solutions and the rational solutions. The advantage of our approach is that we can find the exact solutions for whole Burgers hierarchy. We can construct them
without using the traveling wave. This fact allows us to obtain solutions of different types.

\section{Generalized Cole --- Hopf transformation for solutions of the Burgers hierarchy}

Eq. \eqref{BH} can be
linearized by the Cole---Hopf  transformation \cite{Olver,Polyanin,
Kudryashov92}
\begin{equation}
\label{CH}u=\frac{\Psi_x}{\Psi}, \quad \Psi=\Psi(x,t)
\end{equation}

Taking this transformation into account, we have \cite{Kudryashov92}
\begin{equation}
\label{Eq1}
u_t+\alpha\,\frac{\partial}{\partial x}\left(\frac{\partial}{\partial x}+u\right)^{n}u=\frac{\partial}{\partial
x}\left(\frac{\Psi_t+\alpha\,\Psi_{n+1,x}}{\Psi}\right),
\end{equation}
where $\Psi_{n,x}$ - n-th derivative of $\Psi$ with respect to $x$.

Exact solutions of the Burgers equation can be obtained by using a generalization of the Cole---Hopf transformation \cite{Polyanin, Weiss, Kudryashov88, Kudryashov90a, Kudryashov91,Kudryashov92,Kudryashov09d}. This transformation can be written as
\begin{equation}
\label{Eq2}
u=\frac{F_{x}}{F}+F,\quad F=F(x,t)
\end{equation}
where $F(x,t)$ satisfies the Burgers equation.
Let us show that transformation \eqref{Eq2} is valid for all hierarchy \eqref{BH}.
First of all, we prove the following lemma.

\textbf{Lemma 1}
The following identity takes place
\begin{equation}
\label{L1}
\left(\frac{\partial}{\partial x}+\frac{\Psi_{xx}}{\Psi_{x}}\right)^{n} \frac{\Psi_{xx}}{\Psi_{x}}=\frac{\Psi_{n+2,x}}{\Psi_{x}},
\end{equation}
where $\Psi_{n,x}$ is n-th derivative of $\Psi$ with respect to $x$.

\textbf{Proof.}
Let us apply the method of mathematical induction. When $n=1$ we get
\begin{equation}
\label{L2}
\left(\frac{\partial}{\partial x}+\frac{\Psi_{xx}}{\Psi_{x}}\right) \frac{\Psi_{xx}}{\Psi_{x}}=\frac{\Psi_{xxx}}{\Psi_{x}}
\end{equation}
At $n=2$ we have
\begin{equation}
\label{L3}
\left(\frac{\partial}{\partial x}+\frac{\Psi_{xx}}{\Psi_{x}}\right)^{2} \frac{\Psi_{xx}}{\Psi_{x}}=\left(\frac{\partial}{\partial x}+\frac{\Psi_{xx}}{\Psi_{x}}\right) \frac{\Psi_{xxx}}{\Psi_{x}}=\frac{\Psi_{xxxx}}{\Psi_{x}}
\end{equation}
By the induction, assuming $n=k-1$, we obtain
\begin{equation}
\label{L4}
\left(\frac{\partial}{\partial x}+\frac{\Psi_{xx}}{\Psi_{x}}\right)^{k-1} \frac{\Psi_{xx}}{\Psi_{x}}=\frac{\Psi_{k+1,x}}{\Psi_{x}}
\end{equation}
Finally, when $n=k$ we have
\begin{equation}
\begin{gathered}
\label{L5}
\left(\frac{\partial}{\partial x}+\frac{\Psi_{xx}}{\Psi_{x}}\right)^{k} \frac{\Psi_{xx}}{\Psi_{x}}=\left(\frac{\partial}{\partial x}+\frac{\Psi_{xx}}{\Psi_{x}}\right) \left(\frac{\partial}{\partial x}+\frac{\Psi_{xx}}{\Psi_{x}}\right)^{k-1} \frac{\Psi_{xx}}{\Psi_{x}}=\\\\=
\left(\frac{\partial}{\partial x}+\frac{\Psi_{xx}}{\Psi_{x}}\right) \frac{\Psi_{k+1,x}}{\Psi_{x}}=\frac{\Psi_{k+2,x}}{\Psi_{x}}
\end{gathered}
\end{equation}
This equality completes the proof. \quad
$\Box$

\textbf{Theorem 1}
Let $F(x,t)$ be a solution of Eq. \eqref{BH}.
Then
\begin{equation}
\label{T1}
u=\frac{F_{x}}{F}+F
\end{equation}
is the solution of the Burgers hierarchy  \eqref{BH}.

\textbf{Proof.}
Using the Cole-Hopf transformation \eqref{CH}, we obtain
\begin{equation}
\begin{gathered}
\label{T3}
F_{t}=\frac{\partial}{\partial x}\left(\frac{\Psi_{t}}{\Psi}\right)\\\\
\frac{F_{t}}{F}=\frac{\Psi_{x,t}}{\Psi_{x}}-\frac{\Psi_{t}}{\Psi} \\\\
\frac{F_{x}+F^{2}}{F}=\frac{\Psi_{xx}}{\Psi_{x}}
\end{gathered}
\end{equation}
Substituting transformation \eqref{T1} into hierarchy \eqref{BH} and taking the \textbf{Lemma 1} and Eq. \eqref{T3} into account  we have following set of equalities
\begin{equation}
\begin{gathered}
\label{T2}
\frac{F_{x,\,t}}{F}-\frac{F_{x}F_{t}}{F^{2}}+F_{t}+\alpha \frac{\partial}{\partial x}\left(\frac{\partial}{\partial x}+\frac{F_{x}}{F}+F\right)^{n}
\left(\frac{F_{x}}{F}+F\right)=\\\\
=\frac{\partial}{\partial x} \left(\frac{F_{t}}{F} \right)+F_{t}+\alpha \frac{\partial}{\partial x}\left(\frac{\partial}{\partial x}+\frac{F_{x}+F^{2}}{F}\right)^{n}\left(\frac{F_{x}+F^{2}}{F}\right)=
\end{gathered}
\end{equation}

\begin{equation}
\begin{gathered}
\label{T4}
=\frac{\partial}{\partial x}\left(\frac{\Psi_{x,\,t}}{\Psi_{x}}-\frac{\Psi_{t}}{\Psi}\right)
+\frac{\partial}{\partial x} \left(\frac{\Psi_{t}}{\Psi}\right)
+\alpha \frac{\partial}{\partial x}\left(\frac{\partial}{\partial x}+\frac{\Psi_{xx}}{\Psi_{x}}\right)^{n}\left(\frac{\Psi_{xx}}{\Psi_{x}}\right)=\\\\
=\frac{\partial}{\partial x}\left(\frac{\Psi_{x,\,t}}{\Psi_{x}} + \alpha \left(\frac{\partial}{\partial x}+\frac{\Psi_{xx}}{\Psi_{x}}\right)^{n}\left(\frac{\Psi_{xx}}{\Psi_{x}}\right)\right)=
\end{gathered}
\end{equation}

\begin{equation}
\begin{gathered}
\label{T5}
=\frac{\partial}{\partial x}\left(\frac{\Psi_{x,\,t}}{\Psi_{x}} + \alpha \frac{\Psi_{n+2,x}}{\Psi_{x}}\right)=
\frac{\partial}{\partial x}\left(\frac{1}{\Psi_{x}} \,\frac{\partial}{\partial x} \left(\Psi_{t}+\alpha \Psi_{n+1,x}\right)\right)=0
\end{gathered}
\end{equation}
Thus, we have that if $F(x,t)$ satisfies equation \eqref{BH} then $u(x,t)$ by formula  \eqref{T1}, is solution of \eqref{BH} as well. \quad
$\Box$

\section{Solitary wave solutions of the Burgers hierarchy}

Let us show that the Burgers hierarchy has the
solution in the form
\begin{equation}\begin{gathered}
\label{Sl1}U_l^{(n+1,\,N)}(x,t)=\frac{\sum_{j=1}^{N}\,k_j^l\,
\exp{\left(k_j\,x-\alpha\,k_{j}^{n+1}\,t-x_0^{(j)}\right)}}
{\sum_{j=1}^{N}\,k_j^{l-1}\,\exp{\left(k_j\,x-\alpha\,k_{j}^{n+1}\,t-x_0^{(j)}\right)}},\\
\\
(j=1,2,...N), \qquad (n,l=1,2,...)
\end{gathered}\end{equation}
where $k_j$ and $x_0^{(j)}$ are arbitrary constants.

This result follows from the theorem.

\textbf{Theorem}.  Let
\begin{equation}
\label{Tr}U_0= \frac{\psi_x}{\psi},
\end{equation}
be a solution of the Burgers hierarchy. Then
\begin{equation}
\label{Tr1}U_{k+1}=\frac{\psi_{k+1,x}}{\psi_{k,x}}, \qquad
\psi_{k,x}=\frac{\partial^k\psi}{\partial x^k},
\end{equation}
is a solution of the Burgers hierarchy as well.

\textbf{Proof.} This theorem follows from the generalized
transformation \eqref{T1} for the solution of the
hierarchy \eqref{BH}. Let
\begin{equation}
\label{Tr1a}
U_0=F(x,t)
\end{equation}
be the solution of the hierarchy \eqref{BH} equation, then
\begin{equation}
\label{Tr1b}
U_1=\frac{F_x+F^2}{F}
\end{equation}
is also the solution of the Burgers hierarchy by
the generalized transformation for the solution of the hierarchy \eqref{BH}.

Formulae \eqref{Tr1a} and \eqref{Tr1b} can be written in the form
\begin{equation}
\label{Tr1d}
U_0=\frac{\psi_x}{\psi},\qquad
U_1=\frac{\psi_{xx}}{\psi_{x}},
\end{equation}

Assume that
\begin{equation}
\label{Tr1e}
U_m=\frac{\psi_{m,x}}{\psi_{m-1,x}}, \qquad
\psi_{m,x}=\frac{\partial^m \psi}{\partial x^m},
\end{equation}
is the solution of the hierarchy \eqref{BH} and substituting $U_m$ into the
generalized transformation \eqref{T1} we obtain that
\begin{equation}
\label{Tr1f}
U_{m+1}=\frac{U_{m,x}+U_m^2}{U_m}=\frac{\psi_{m+1,x}}{\psi_{m,x}}
\end{equation}
is a solution of hierarchy \eqref{BH}.

This equality completes the proof. \quad
$\Box$

This theorem allows us to have the solutions of the
Burgers hierarchy in the form \eqref{Sl1}.

It is obvious, that function
\begin{equation}
\label{E1}\psi^{n+1}(x,t)=\sum_{j=0}^{N}\exp{\left(\,z_j\right)},\qquad
z_j=k_j\,x-\alpha\,k_{j}^{n+1}\,t-x_0^{(j)}
\end{equation}
is the solution of the
\begin{equation}
\label{Lin}\psi_{t}+\alpha \, \psi_{n+1,\,x}=0
\end{equation}

By the Cole --- Hopf transformation \eqref{CH} we have the solution of
the Burgers hierarchy in the form
\begin{equation}
\label{S2}U^{(n+1,\,N)}=\frac{\sum_{j=0}^{N}k_j\,\exp{\left(\,z_j\right)},}
{\sum_{j=0}^{N}\exp{\left(\,z_j\right)},}\qquad
z_j=k_j\,x-\alpha\,k_{j}^{n+1}\,t-x_0^{(j)}
\end{equation}

Taking the theorem into account we have the solution of the hierarchy \eqref{BH}
in the form \eqref{Sl1}.

Let us present some examples.
When $N=2$, $n=l=1$ we have the solution of
the Burgers equation
\begin{equation}\begin{gathered}
\label{S0_0}U_1^{(2,2)}=\frac{k_1\,\exp{\left(z_1\right)}+k_2\,\exp{\left(z_2\right)}}
{\exp{\left(z_1\right)}+\exp{\left(z_2\right)}},\\\\
z_j=k_jx-\alpha\,k_{j}^2\,t-x_0^{(j)},\quad (j=1,2)
\end{gathered}\end{equation}
In the case $N=2$, $n=l=2$ we have the solution of the Sharma---Tasso---Olver equation in the form
\begin{equation}\begin{gathered}
\label{S1}U_2^{(3,2)}=\frac{k_1^2\,\exp{\left(z_1\right)}+k_2^2\,\exp{\left(z_2\right)}}
{k_1\exp{\left(z_1\right)}+k_2\exp{\left(z_2\right)}},\\\\
z_j=k_jx-\alpha\,k_{j}^3\,t-x_0^{(j)},\quad (j=1,2)
\end{gathered}\end{equation}

When $N=3$, $n=2$  and $l=5$ we obtain the following solution of the
Sharma---Tasso---Olver equation
\begin{equation}\begin{gathered}
\label{S23}U_5^{(3,3)}=\frac{k_1^5\,\exp{\left(z_1\right)}+k_2^5\,\exp{\left(z_2\right)}+
k_3^5\,\exp{\left(z_3\right)}}
{k_1^{4}\exp{\left(z_1\right)}+k_2^{4}\exp{\left(z_2\right)}+
k_3^{4}\,\exp{\left(z_3\right)}},\\
\\
z_j=k_jx-\alpha\,k_{j}^3\,t-x_0^{(j)},\quad (j=1,2,3)
\end{gathered}\end{equation}

In the case $N=2$, $n=3$ and $l=3$ we have solitary wave solution for the Eq. \eqref{FM}
\begin{equation}\begin{gathered}
\label{S32}U_2^{(4,2)}=\frac{k_1^3\,\exp{\left(z_1\right)}+k_2^3\,\exp{\left(z_2\right)}}
{k_1^2\exp{\left(z_1\right)}+k_2^2\exp{\left(z_2\right)}},\\
\\
z_j=k_jx-\alpha\,k_{j}^4\,t-x_0^{(j)},\quad (j=1,2)
\end{gathered}\end{equation}

We can see that Eq. \eqref{Lin} is linear and has polynomial solutions. Thus, we can
present its solution in the form
\begin{equation}\begin{gathered}
\label{S24}
\Psi^{n+1}(x,t)=\sum_{i=0}^{I} C_{i}\,x^{i}+\sum_{j=0}^{N} e^{z_{j}},\\\\
z_j=k_jx-\alpha\,k_{j}^{n+1}\,t-x_0^{(j)},\\\\
(j=0,1,2,\ldots N), \qquad (n=1,2,\ldots),\qquad N \in\mathbb{N},\\\\
(I=0,1,2,\ldots,n), \qquad (i=0,1,2,\ldots I).
\end{gathered}\end{equation}
\noindent
From the transformation \eqref{CH} and formula \eqref{S24} we have the exact
solution of the hierarchy \eqref{BH} in the form
\begin{equation}\begin{gathered}
\label{S25} U^{(n+1,\,N)}(x,t)={\frac {\sum_{i=0}^{I} i C_{i} \,x^{i-1}+ \sum_{j=0}^{N} k_{j} e^{z_{j}}}{\sum_{i=0}^{I} C_{i}\,x^{i}+ \sum_{j=0}^{N} e^{z_{j}}}} ,\\\\
(j=0,1,2,\ldots N), \qquad (n=1,2,\ldots),\qquad N \in\mathbb{N},\\\\
(I=0,1,2,\ldots,n), \qquad (i=0,1,2,\ldots I).
\end{gathered}\end{equation}

For the Burgers equation (n=1) from \eqref{S25} we obtain the solution in the form
\begin{equation}\begin{gathered}
\label{Sl2}
U^{(2,2)}(x,t)={\frac {C_{1}+k_{1}{{\rm e}^{k_{1}x-\alpha\,k_{1}^{2}t-x_1}}+k_{{2}}{{\rm e}^{k_{{2}}x-\alpha\,k_{2}^{2}t- x_2}}}{C_{0}+C_{1}x+{{\rm e}^{k_{1}x-\alpha\,k_{1}^{2}t- x_1}}+{
{\rm e}^{k_{2}x-\alpha\,k_{2}^{2}t- x_2}}}}
\end{gathered}\end{equation}
We demonstrate solution \eqref{Sl2} when $C_{0}=C_{1}=k_{1}=1,\,k_{2}=2$ on Fig. 1.
For the Eq. \eqref{FM} from \eqref{S25} we have the following solution
\begin{equation}\begin{gathered}
\label{Sl2}
U^{(4,1)}(x,t)=\frac {C_{1}+2C_{2}x+3C_{3}x^{2}+k_{1}{{\rm e}^{k_{{1}}x-\alpha\,k_{1}^{4}t-
x_1}}}{C_{0}+C_{1}x+C_{2}x^{2}+C_{3}x^{3}+{\rm e}^{k_{1}x-\alpha\,k_{1}^{4}t- x_1}}
\end{gathered}\end{equation}
\begin{figure}[h]
\centering
 \includegraphics[width=70mm]{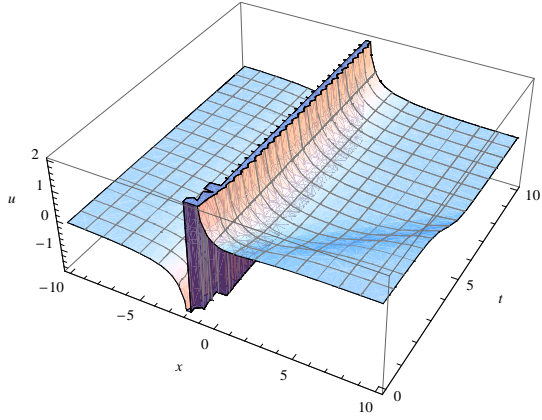}
 \caption{The solution \eqref{Sl2} of the Burgers equation}
 \label{fig1}
\end{figure}

By analogy with solution \eqref{Sl1} we can look for the periodic
solutions of the equation for the Burgers hierarchy taking the trigonometric functions
into consideration.  Equation \eqref{Lin} has trigonometric solutions at $n+1=2l+1$
in the form
\begin{equation}\begin{gathered}
\label{S25}
\Psi^{2l+1}(x,t)=\sum_{i=0}^{I} C_{i}\,x^{i}+\sum_{m=0}^{M} \sin  z^{'}_{m} + \sum_{p=0}^{P} \cos z^{'}_{p}  +\sum_{j=0}^{N} e^{z_{j}},\\\\
z_j=k_{j}\,x-\alpha\,k_{j}^{2l+1}\,t-x_{0}^{j},\\\\
z^{'}_{m,p}=k_{m,p}^{(1,2)}\,x+(-1)^{l+1}\,\alpha\,(k_{{m,p}}^{(1,2)})^{2l+1}\,t-x_{(1,2)}^{(m,p)},\\\\
(j=0,1,2,\ldots,N), \qquad (l=1,2,\ldots), \quad N \in \mathbb{N}, \\\\
(I=0,1,2,\ldots,2l), \qquad (i=0,1,2,\ldots I),\\\\
(m,p=0,1,2,\ldots,M), \qquad M,\,P \in \mathbb{N}.
\end{gathered}\end{equation}

From the transformation \eqref{CH} we have  the exact
solution of the hierarchy in the form \eqref{BH}
\begin{equation}\begin{gathered}
\label{S26} U^{(2l+1)}(x,t)=\\\\=\frac {\sum_{i=0}^{I} i C_{i} \,x^{i-1}+\sum_{m=0}^{M} k_{m}^{(1)} \cos  z^{'}_{m}-\sum_{p=0}^{P}\,k_{p}^{(2)} \sin z^{'}_{p}+ \sum_{j=0}^{N} k_{j} e^{z_{j}}}{\sum_{i=0}^{I} C_{i}\,x^{i}+\sum_{m=0}^{M} \sin  z^{'}_{m} + \sum_{p=0}^{P} \cos z^{'}_{p}  +\sum_{j=0}^{N} e^{z_{j}}}
\end{gathered}\end{equation}

For example, we can write following solution for the Sharma -Tasso - Olver equation (l=1)
\begin{equation}\begin{gathered}
\label{Sl3} U^{(3)}(x,t)=\frac {k_1{e^{k_1 x-\alpha
k_1^{3}t-x_1}}+ \cos \left( k_2 x+\alpha k_2^{3}t-x_2
\right) k_2-\sin \left( k_3 x+\alpha k_3^{3}t-x_3 \right)
k_3}{C_0+e^{k_1x-\alpha k_1^{3}t-x_1}+ \sin \left(
{k_2}x+\alpha k_2^{3}t-{x_2} \right) +\cos
 \left( k_3 x+\alpha k_3^{3}t-{x_3} \right) }
\end{gathered}\end{equation}
Assuming $C_0=8$, $k_1=2.2$, $k_2=3$, $k_3=6$, $x_1=0.3$, $x_2=6$ and $x_3=2$ we demonstrate solution \eqref{Sl3} on Fig. 2.
\begin{figure}[h]
\centering
 \includegraphics[width=70mm]{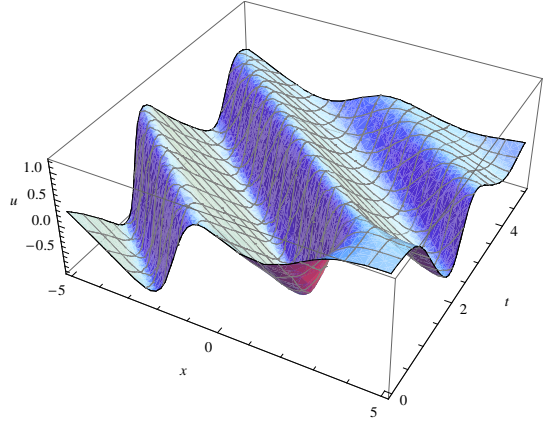}
 \caption{The solution \eqref{Sl3} of STO equation}
 \label{fig2}
\end{figure}

From formula \eqref{S26} we obtain following solution for the Eq. \eqref{FFM} (l=2)
\begin{equation}\begin{gathered}
\label{Sl3} U^{(5)}(x,t)=\frac {C_{0}+k_1 \cos \left( k_1 x+\alpha k_1^{5}t-x_1
\right) -k_2 \sin \left( k_2 x+\alpha k_2^{5}t-x_2 \right)
}{C_0+C_{1}x+\sin \left(k_1x+\alpha k_1^{5}t-x_1 \right) +\cos\left( k_2 x+\alpha k_2^{5}t-x_2 \right) }
\end{gathered}\end{equation}

Other solutions can be written using the formula \eqref{Tr1}.

\section{Rational solutions of the Burgers hierarchy}

Using Eq. \eqref{Lin} and transformations \eqref{Tr1} and
\eqref{Tr1a} we can find the rational solutions of the hierarchy \eqref{BH}.
To obtain these solutions we use the
solutions of Eq. \eqref{Lin} in the form
\begin{equation}\begin{gathered}
\label{R1}\psi_0(x,t)=1,\quad \psi_1(x,t)=x, \ldots, \psi_{n}(x,t)=x^{n}
\end{gathered}\end{equation}

Integrating $\psi_{n}(x,t)=x$ with respect to $x$ we obtain
$\psi_{n+1}(x,t)=x^2+\varphi_{n+1}(t)$. Substituting $\psi_{n+1}(x,t)$ into Eq.
\eqref{Lin} we get $\varphi_{n+1}=-(n+1)!\,\alpha\,t$.  Substituting
$\psi_{n+1}(x,t)$ into Eq. \eqref{Lin} we obtain
\begin{equation}\begin{gathered}
\label{R1a}\quad\psi_{n+1}(x,t)=x^{n+1}-(n+1)!\,\alpha\,t.
\end{gathered}\end{equation}

Continuing in the same way, we can obtain the solutions $\psi_q(x,t), q=n+2,n+3,\ldots$ as
a result of integration of solution with respect to $x$. Taking these polynomial solutions of \eqref{Lin} into account we obtain the rational solutions of the Burgers hierarchy \eqref{BH}.

The polynomial solutions of \eqref{Lin} for $n=2$  are the following
\begin{equation}\begin{gathered}
\label{R0}
\psi_0(x,t)=1,\quad \psi_1(x,t)=x,\quad \psi_2(x,t)=x^{2},
\end{gathered}\end{equation}

\begin{equation}\begin{gathered}
\label{R1}
\psi_3(x,t)=x^{3}-6\,\alpha\,t,
\end{gathered}\end{equation}

\begin{equation}\begin{gathered}
\label{R2}\psi_4(x,t)=x^4-24\,\alpha\,x\,t,
\end{gathered}\end{equation}

\begin{equation}\begin{gathered}
\label{R3}\psi_5(x,t)={x}^{5}-60\,\alpha\,{x}^{2}t,
\end{gathered}\end{equation}

\begin{equation}\begin{gathered}
\label{R4}\psi_6(x,t)={x}^{6}-120\,\alpha\,{x}^{3}t+360\,{\alpha}^{2}{t}^{2},
\end{gathered}\end{equation}

\begin{equation}\begin{gathered}
\label{R5}\psi_7(x,t)={x}^{7}-210\,\alpha\,{x}^{4}t+2520\,{\alpha}^{2}x{t}^{2},
\end{gathered}\end{equation}

\begin{equation}\begin{gathered}
\label{R6}\psi_8(x,t)={x}^{8}-336\,\alpha\,{x}^{5}t+10080\,{\alpha}^{2}{x}^{2}{t}^{2},
\end{gathered}\end{equation}

\begin{equation}\begin{gathered}
\label{R7}\psi_9(x,t)={x}^{9}-504\,\alpha\,{x}^{6}t+30240\,{\alpha}^{2}{x}^{3}{t}^{2}-60480
\,{\alpha}^{3}{t}^{3},
\end{gathered}\end{equation}

\begin{equation}\begin{gathered}
\label{R8}\psi_
{10}(x,t)={x}^{10}-720\,\alpha\,{x}^{7}t+75600\,{x}^{4}{\alpha}^{2}{t}^{2}-
604800\,{\alpha}^{3}x{t}^{3},
\end{gathered}\end{equation}

\begin{equation}\begin{gathered}
\label{R9}\psi_{11}(x,t)={x}^{11}-990\,\alpha\,{x}^{8}t+166320\,{x}^{5}{\alpha}^{2}{t}^{2}-
3326400\,{\alpha}^{3}{t}^{3}{x}^{2},
\end{gathered}\end{equation}

\begin{equation}\begin{gathered}
\label{R10}\psi_{12}(x,t)={x}^{12}-1320\,\alpha\,{x}^{9}t+332640\,{x}^{6}{\alpha}^{2}{t}^{2}-
\\-
13305600\,{\alpha}^{3}{t}^{3}{x}^{3}+19958400\,{\alpha}^{4}{t}^{4},
\end{gathered}\end{equation}

Taking into account these solutions we have the rational solutions
of the Sharmo---Tasso---Olver equation in the form
\begin{equation}\begin{gathered}
\label{ss0}U_1(x,t)=\frac {1}{x}, \qquad U_2(x,t)=\frac{2}{x}
\end{gathered}\end{equation}

\begin{equation}\begin{gathered}
\label{ss1}U_3(x,t)=3\,{\frac {{x}^{2}}{{x}^{3}-6\,\alpha\,t}},
\end{gathered}\end{equation}

\begin{equation}\begin{gathered}
\label{ss2}U_4(x,t)=4\,{\frac {{x}^{3}-6\,\alpha\,t}{x \left(
{x}^{3}-24\,\alpha\,t \right) }},
\end{gathered}\end{equation}

\begin{equation}\begin{gathered}
\label{ss3}U_5(x,t)=5\,{\frac {{x}^{3}-24\,\alpha\,t}{x \left(
{x}^{3}-60\,\alpha\,t \right) }},
\end{gathered}\end{equation}

\begin{equation}\begin{gathered}
\label{ss4}U_6(x,t)=6\,{\frac {{x}^{2} \left( {x}^{3}-60\,\alpha\,t
\right) }{{x}^{6}-120 \,\alpha\,{x}^{3}t+360\,{\alpha}^{2}{t}^{2}}},
\end{gathered}\end{equation}

\begin{equation}\begin{gathered}
\label{ss5}U_7(x,t)=7\,{\frac
{{x}^{6}-120\,\alpha\,{x}^{3}t+360\,{\alpha}^{2}{t}^{2}}{x \left(
{x}^{6}-210\,\alpha\,{x}^{3}t+2520\,{\alpha}^{2}{t}^{2} \right) }},
\end{gathered}\end{equation}

\begin{equation}\begin{gathered}
\label{ss6}U_8(x,t)=8\,{\frac
{{x}^{6}-210\,\alpha\,{x}^{3}t+2520\,{\alpha}^{2}{t}^{2}}{x \left(
{x}^{6}-336\,\alpha\,{x}^{3}t+10080\,{\alpha}^{2}{t}^{2} \right) }},
\end{gathered}\end{equation}

\begin{equation}\begin{gathered}
\label{ss7}U_9(x,t)=9\,{\frac {{x}^{2} \left(
{x}^{6}-336\,\alpha\,{x}^{3}t+10080\,{\alpha }^{2}{t}^{2} \right)
}{{x}^{9}-504\,\alpha\,{x}^{6}t+30240\,{\alpha}^{
2}{x}^{3}{t}^{2}-60480\,{\alpha}^{3}{t}^{3}}},
\end{gathered}\end{equation}

\begin{equation}\begin{gathered}
\label{ss8}U_{10}(x,t)=10\,{\frac
{{x}^{9}-504\,\alpha\,{x}^{6}t+30240\,{\alpha}^{2}{x}^{3}{t
}^{2}-60480\,{\alpha}^{3}{t}^{3}}{x \left(
{x}^{9}-720\,\alpha\,{x}^{6
}t+75600\,{\alpha}^{2}{x}^{3}{t}^{2}-604800\,{\alpha}^{3}{t}^{3}
\right) }},
\end{gathered}\end{equation}

\begin{equation}\begin{gathered}
\label{ss9}U_{11}(x,t)=11\,{\frac
{{x}^{9}-720\,\alpha\,{x}^{6}t+75600\,{\alpha}^{2}{x}^{3}{t
}^{2}-604800\,{\alpha}^{3}{t}^{3}}{x \left(
{x}^{9}-990\,\alpha\,{x}^{
6}t+166320\,{\alpha}^{2}{x}^{3}{t}^{2}-3326400\,{\alpha}^{3}{t}^{3}
 \right) }}.
\end{gathered}\end{equation}

\begin{figure}[h]
\centering
 \includegraphics[width=70mm]{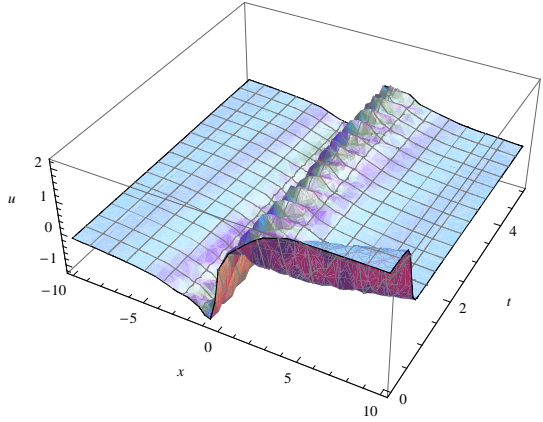}
 \caption{The solution \eqref{ss10} of the STO equation}
 \label{fig2}
\end{figure}

Using the solution of Eq.\eqref{Lin} as the sum of rational,
exponential functions and, at $n=2l$, trigonometric functions we can obtain many solutions of the
hierarchy \eqref{BH}. In particulare, at $n=2$, taking into account
solution in the form
\begin{equation}\begin{gathered}
\label{ss10}\Psi(x,t)={C_2}\, \left( 1+{x}^{2} \right) +{e^{{
k_1}\,x-\alpha\,{{k_1} }^{3}t-{x_1}}}+\cos \left( {
k_2}\,x+\alpha\,{{k_2}}^{3}t-{x_2} \right)
\end{gathered}\end{equation}
we have solution of the Sharma---Tasso---Olver equation in the form
\begin{equation}\begin{gathered}
\label{ss10}U(x,t)=\frac{2\,{C_2}\,x+{k_1}\,{e^{{
k_1}\,x-\alpha\,{{k_1}}^{3}t-{x_1}}}-\sin \left( {
k_2}\,x+\alpha\,{{k_2}}^{3}t-{x_2}
 \right) {k_3}}{{C_2}\, \left( 1+{x}^{2} \right)
+{e^{{k_1}\,x-\alpha\,{{k_1} }^{3}t-{x_1}}}+\cos \left( {
k_2}\,x+\alpha\,{{k_2}}^{3}t-{x_2} \right) }
\end{gathered}\end{equation}

Assuming $C_2=8$, $k_1=2.2$, $k_2=6$, $x_1=0.3$, and $x_2=2$ we
obtain solution \eqref{ss10} on Fig. 3.

\section{Conclusion}
In this paper the generalized Cole-Hopf transformation  was found for the Burgers hierarchy.
We have presented some classes of the exact solutions for the
Burgers hierarchy. These classes are expressed via the
rational, exponential and triangular functions and as the sum of
these functions.


\begin{thebibliography}{99}



\bibitem{Burgers} J.M. Burgers, A mathematical model illustrating the theory of turbulance, Advances in Applied Mechanics.1 (1948) 171-199.

\bibitem{Hopf51} E. Hopf, The partial differential equation $u_t+u\,u_x=u_{xx}$,
Communs. Pure Appl. Math. 3 (1950) 201-230.

\bibitem{Cole50} J.D. Cole, On a quasi-linear parabolic equation occuring in aerodynamics
 Quart. Appl. Math. 9 (1950) 225-236.

\bibitem{Rosenblatt68}  M. Rosenblatt, Remark on the Burgers equation,
 Phys. Fluids. 9 (1966) 1247-1248.

\bibitem{Benton66}  E.R. Benton,  Some New Exact, Viscous, Nonsteady Solutions of Burgers' Equation,
 J. Math. Phys.  9 (1968) 1129-1136.

\bibitem{Malfliet} W. Malfliet, Approximate solution of the damped Burgers equation, J. Phys. A. 26 (1993) L723-L728.

\bibitem{Fahmy} E. S. Fahmy, K. R. Raslan, H. A. Abdusalam, On the exact and numerical solution of the time-delayed Burgers equation, International Journal of Computer Mathematics. 85 (2008) 1637-1648

\bibitem{Tasso} A. S. Sharma , H. Tasso,  Connection between wave envelope and explicit solution of a nonlinear
dispersive equation. Report IPP 6/158. 1977.

\bibitem{Olver} P.J. Olver,  Evolution equations possessing infinitly many symmetries, J. Math. Phys. 18 (1977) 1212-1215.

\bibitem{Hereman} W. Heremant, P.P. Banerjee, A. Korpel, G. Assanto,
A. Van Immerzeele, A. Meerpoel, Exact solitary wave solutions of non-linear evolution and wave
equations using a direct algebraic method, J. Phys. A Math. Gen. 19 (1986) 607-628.

\bibitem{Yang} Z. J. Yang, Travelling wave solutions to nonlinear evolution and wave
equations, J. Phys. A Math. Gen. 27 (1994) 2837-2855.

\bibitem{Gudkov} V.V. Gudkov,  A family of exact travelling wave solutions to nonlinear
evolution and wave equations, J. Math. Phys. 38 (1997) 4794-4803.

\bibitem{Wazwaz} A.-M. Wazwaz, New solitons and kinks solutions to the Sharma–--Tasso–--Olver equation, Applied Mathematics and Computation. 188 (2007) 1205-1213.

\bibitem{Shang} Y. Shanga, J. Qina, Y. Huangb, W. Yuana, Abundant exact and explicit solitary wave and periodic wave solutions to the Sharma–--Tasso–--Olver equation, Applied Mathematics and Computation. 202 (2008) 532-538.

\bibitem{Lou} S. Wang, X. Tang, S.-Y. Lou, Soliton fission and fusion: Burgers equation and
Sharma--–Tasso--–Olver equation. Chaos, Solitons and Fractals. 21 (2004) 231-239.

\bibitem{Kudryashov08} N.A. Kudryashov, N.B. Loguinova  Extended simplest equation method for nonlinear differential equations , Applied Mathematics and Computation. 205 (2008) 396 - 402.

\bibitem{Polyanin} A.D. Polyanin, V.F.  Zaitsev, A.I. Zhyrov, Methods
of nonlinear equations of mathematical physics and mechanics,
Fizmatlit, Moscow, 2005.

\bibitem{Kudryashov92} N.A. Kudryashov,  Partial differential equations with
solutions having movable forst - order singularities, Physics
Letters A. 169 (1992) 237 - 242.

\bibitem{Weiss}  J. Weiss,  M. Tabor , G. Carnevalle, The Painleve property
for nonlinear partial differential equations, J. Math Phys.
24 (1983) 522.

\bibitem{Kudryashov88} N.A. Kudryashov Exact soliton solutions of the generalized
evolution equation of wave dynamics, Journal of Applied Mathematics
and Mechanics. 52 (1988) 360--365.

\bibitem{Kudryashov90a} N.A. Kudryashov Exact soliton solutions of nonlinear wave
equations arising in mechanics, Journal of Applied Mathematics and
Mechanics. 54 (1990) 450 - 453.

\bibitem{Kudryashov91} N.A. Kudryashov On types of nonlinear nonintegrable equations
with exact solutions Phys Lett. A. 155 (1991) 269 - 275.

\bibitem{Kudryashov09d} N.A. Kudryashov, N.B. Loguinova  Be carefull with Exp - function
method, Commun Nonlinear Sci Numer Simulat. 14 (2009) 1881 - 1890.


\end{thebibliography}
\end{document}